\documentclass[
final]{aipproc}

\layoutstyle{8x11double}
\usepackage{graphicx,epstopdf}
\usepackage{epsfig}
\usepackage{pst-all}
\usepackage{float}
\usepackage{amsmath}
\usepackage{amssymb}
\usepackage{tikz}
\usetikzlibrary{fit,backgrounds,positioning,decorations.markings,fadings,patterns,decorations.pathmorphing,arrows}

\begin{document}

\title{Non-minimal Higgs Inflation and Frame Dependence in Cosmology}

\classification{\texttt{11.10.Hi; 14.80.Bn; 98.80.Cq; 04.60.-m; 98.80.Qc; 11.10.Gh}}

\author{Christian F. Steinwachs}{
  address={School of Mathematical Sciences, University of Nottingham
University Park, Nottingham, NG7 2RD, UK}}

\author{Alexander Yu. Kamenshchik}{
  address={Dipartimento di Fisica e Astronomia and INFN, Via Irnerio 46, 40126 Bologna, Italy}, altaddress={L.D. Landau Institute for Theoretical Physics of the Russian Academy of Sciences, Kosygin str. 2, 119334 Moscow, Russia} }

\begin{abstract}
We investigate a very general class of cosmological models with scalar fields non-minimally coupled to gravity. A particular representative in this class is given by the non-minimal Higgs inflation model in which the Standard Model Higgs boson and the inflaton are described by one and the same scalar particle. While the predictions of the non-minimal Higgs inflation scenario come numerically remarkably close to the recently discovered mass of the Higgs boson, there remains a conceptual problem in this model that is associated with the choice of the cosmological frame. While the classical theory is independent of this choice, we find by an explicit calculation that already the first quantum corrections induce a frame dependence. We give a geometrical explanation of this frame dependence by embedding it into a more general field theoretical context. From this analysis, some conceptional points in the long lasting cosmological debate: ``Jordan frame vs. Einstein frame'' become more transparent and in principle can be resolved in a natural way.
\end{abstract}

\maketitle

\section{Class of cosmological models}

Let us consider the action of a very general class of cosmological models with a non-minimal coupling
\begin{align}
S=\int\text{d}^4x\,\sqrt{g}\left(U(\varphi)\,R-\frac{G(\varphi)}{2}\,\nabla_{\mu}\varPhi^{a}\nabla^{\mu}\varPhi_{a}-V(\varphi)\right)\;.\label{ONAction}
\end{align}
The multiplet of scalar fields $\varPhi^{a},\;a=1,...,N$ has a rigid internal $O(N)$ symmetry. Internal indices are raised and lowered by the constant metric $\delta_{ab}$. Representatives of this class are parametrized by different choices of the functions $U(\varphi)$, $G(\varphi)$ and $V(\varphi)$ and the number of scalar components $N$. In order not to spoil the internal $O(N)$ symmetry, $U(\varphi)$, $G(\varphi)$ and $V(\varphi)$ must be ultra-local functions of the modulus $\varphi:=\sqrt{\varPhi^{a}\varPhi_{a}}$. In \cite{OneLoopON}, the divergent part of the one-loop effective action for this class of models has been calculated in a closed form. Due to its universality, (\ref{ONAction}) covers many important cosmological models, so that the results of \cite{OneLoopON} immediately yield the one-loop corrections to the desired model. In the following, we will make use of these results several times.

\section{Non-minimal Higgs Inflation}
\begin{figure}
\includegraphics[scale=0.9]{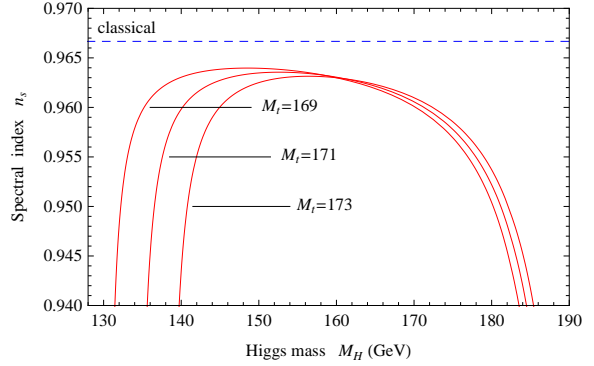}
\caption{Running spectral index $n_{\text{s}}$ as a function of the Higgs mass $M_{\text{H}}$ for different values of the top-quark mass $M_{\text{t}}$.}
\label{SpectralIndex}
\end{figure}

The basic idea of Higgs inflation is that the Standard Model (SM) Higgs boson
and the inflaton are one and the same scalar particle. In order to ensure consistency with observational data, it is necessary to assume a non-minimal coupling $\propto\xi\,\varphi^2\,R$ to gravity with a large coupling constant $\xi\simeq 10^4$.
The tree-level graviton-Higgs sector of the non-minimal Higgs inflation model is described by (\ref{ONAction}) for the choices
\begin{align}
U_{\text{tree}}&{}=\frac{1}{2}\,(M_{\text{P}}^2+\xi\,\varphi^2),\quad G_{\text{tree}}=1,\nonumber\\
V_{\text{tree}}&{}=\frac{\lambda}{4}(\varphi^2-\nu^2)^2,\quad N=4\;.\label{TreeLevelNMHCouplings}
\end{align}
The matter sector of the model is given by the interaction part of the SM and can be described schematically as
\begin{align}
{\cal L}_{\text{int}}^{\text{SM}}=-\sum_{\chi}\frac{\lambda_{\chi}}{2}\chi^2\varphi^2-\sum_{A}\frac{g_{A}^2}{2}A_{\mu}^2\varphi^2-\sum_{\Psi}y_{\Psi}\varphi\,\bar{\Psi}\,\Psi
\end{align}
with the sums extending over scalar fields $\chi$, vector gauge fields $A_{\mu}$ and Dirac spinors $\Psi$. The $\lambda_{\chi}$, $g_{A}$ and $y_{\Psi}$ are the corresponding scalar, gauge and Yukawa couplings. This sector is dominated by the heavy masses
\begin{align}
m^2_{W^{\pm}}=\frac{g^2}{4}\varphi^2,\;\; m_{Z}^2=\frac{(g^2+g'^2)}{4}\varphi^2,\;\; m_{\text{t}}^2=\frac{y_{\text{t}}^2}{2}\varphi^2\label{Masses}
\end{align}
of the $W^{\pm}$ and $Z$ bosons and the Yukawa top-quark $q_{\text{t}}$.
A key feature of this model is a suppression mechanism that is induced by the non-minimal coupling and can be parametrized by the suppression function
\begin{align}
s(\varphi):=\frac{U^2}{G\,U+3\,U'^2}=\frac{M_{\text{P}}^2+\xi\varphi^2}{M_{\text{P}}^2+(6\,\xi+1)\,\xi\,\varphi^2}\label{SuppressionFunction}
\end{align}
with ``prime'' meaning differentiation with respect to $\varphi$. For high energies $\varphi\gg M_{\text{P}}/\sqrt{\xi}$, the suppression function (\ref{SuppressionFunction}) behaves as $s\simeq1/(6\xi)$. The predictions derived in this model depend sensitively on the quantum corrections. Using the one-loop results of \cite{OneLoopON} with the choices (\ref{TreeLevelNMHCouplings}), the one-loop corrections to (\ref{TreeLevelNMHCouplings}) are given by \cite{BarKamStar}
\begin{align}
U_{1-\text{loop}}&{}=\frac{\varphi^2}{32\,\pi^2}\,\mathbf{C}\,\text{ln}\,\frac{\varphi^2}{\mu^2},\quad G_{1-\text{loop}}=\frac{1}{32\,\pi^2}\,\mathbf{E}\,\text{ln}\,\frac{\varphi^2}{\mu^2},\nonumber\\
V_{1-\text{loop}}&{}=\frac{\lambda\,\varphi^4}{32\,\pi^2}\,\mathbf{A}\,\text{ln}\,\frac{\varphi^2}{\mu^2}\label{V1loop}
\end{align}
It is important to note that only the propagator of the radial Higgs mode is suppressed by (\ref{SuppressionFunction}). The angular Goldstone modes that run in the loop do not feel this suppression. For high energies, the logarithmic prefactors $C$, $E$ and $A$ therefore only receive Goldstone but no Higgs contributions. Neglecting graviton loops\footnote{These are obviously suppressed by the effective Planck mass $M_{\text{P}}^{\text{eff}}:=\sqrt{M_{\text{P}}^2+\xi\,\varphi^2}\gg M_{\text{P}}$} and expanding in inverse powers of $\xi$, leads to
\begin{align}
&{}\mathbf{A}=\frac{3}{8\,\lambda}\left(2\,g^4+(g^2+g'^2)^2-16\,y_{\text{t}}^4\right)+6\,\lambda+{\cal O}(\xi^{-2}),\nonumber\\
&{}\mathbf{C}= 3\,\xi\,\lambda+{\cal O}(\xi^0),\quad
\mathbf{E}=0+{\cal O}(\xi^{-2})\,.
\end{align}
In the expression for the anomalous scaling $\mathbf{A}$, the contribution of ordinary matter (\ref{Masses}) to the effective potential has also been included.
In order to establish contact to the usual slow-roll formalism of inflation, we perform a transformation from the original field variables - denoted Jordan frame (JF) parametrization - to the so called Einstein frame (EF). The name EF derives from the fact that expressed in this field variables, the action formally resembles the situation of General Relativity (GR) minimally coupled to a scalar field.\footnote{But it is of course \emph{not} equivalent to a scalar field minimally coupled to GR. One easy way to see this is to notice that the equivalence principle is violated in the JF due to the non-minimal coupling. Since the equivalence principle is a \emph{physical} principle (it does not ``know'' anything about a parametrization at all), it is of course still violated in the EF since we still describe the same theory.} The field transformation from the JF to the EF involves a conformal transformation of the metric field, a non-linear transformation of the scalar field and a re-definition of the potential (EF quantities are denoted by a hat)
\begin{align}
\hat{g}_{\mu\nu}&{}=\frac{2\,U}{M_{\text{P}}^{2}}\,g_{\mu\nu},\quad\left(\frac{\text{d}\hat{\varphi}}{\text{d}\varphi}\right)^2=\frac{M_{\text{P}}^2}{2}\,\frac{G\,U+3\,(U')^2}{U^2},\nonumber\\
 \hat{V}(\hat{\varphi})&{}=\left.\left(\frac{M_{\text{P}}^2}{2}\right)^2\,\frac{V(\varphi)}{U^2(\varphi)}\right|_{\varphi=\varphi(\hat{\varphi})}\,.\label{JFtoEF}
\end{align}
All cosmological parameters can be expressed in terms of the EF effective potential and derivatives thereof
\begin{align}
\hat{V}_{\text{eff}}(\hat{\varphi})\simeq\frac{\lambda M_{\text{P}}^4}{4\xi^2}\;\left(1-\frac{2M_{\text{P}}^2 }{\xi\varphi^2}+\frac{\mathbf{A_{\text{I}}}}{16\pi^2}\ln{\frac{\varphi}{\mu}}\right)\,.
\end{align}
Moreover, it turns out that their behaviour is essentially determined by one single quantity - the inflationary anomalous scaling $\mathbf{A}_{\text{I}}:=A-12\,\lambda$. The extra term $\propto-12\,\lambda$ is due to the quantum Goldstone contributions contained in $\mathbf{C}$. However, calculating the quantum corrections is not sufficient. Since we intend to connect two energy scale that are separated by many orders of magnitude, we have to take into account the dependence of the coupling constants on the energy scale. Thus, we must calculate the beta functions and evaluate the Renormalization Group (RG) flow of the couplings from the electroweak scale $\varphi\simeq\nu$ up to the high energies $\varphi\gg M_{\text{P}}/\sqrt{\xi}$ during inflation \cite{BezShap1,Wil, weAsympt}. The beta functions form a complicated system of coupled ordinary differential equations that can only be solved numerically \cite{weAsympt}. Once this is done, we can express all cosmological parameters in terms of these couplings evaluated at inflationary energy scales. The most important result is the spectral index as a function of the Higgs mass - shown in Fig. \ref{SpectralIndex} - that leads to a constraint of the Higgs mass \cite{weAsympt}
\begin{align}
135.6\; \text{GeV} \lesssim M_{\text{H}}\lesssim 184.5\;\text{GeV}\;.\label{HiggsMassRange}
\end{align}
\noindent
\noindent The announcement of the LHC suggests a Higgs mass around $M_{\text{H}}\simeq 126$ GeV. Although the allowed window (\ref{HiggsMassRange}) does not contain this value, it still comes remarkably close to it.\footnote{This numerical result is highly non-trivial, as a priori, it was by no means guaranteed that the model would predict reasonable values for $M_{\text{H}}$. Only the combination of many features including the RG running leads to values of $M_{\text{H}}$ close to the measured one.}
Moreover, the numerical result can be improved by including higher loops and ultimately leads to a shift of about $10$ GeV bringing the lower bound of $M_{\text{H}}$ numerically very close to the measured value of $M_{\text{H}}$, see e.g. \cite{BezShap2}.\footnote{One should keep in mind that, independently of the higher order quantum corrections, the result also depends sensitively on the initial conditions at the electroweak scale. This uncertainty in the parameter space was not yet fully exploited and can also shift the lower bound towards lower values of $M_{\text{H}}$. As can be seen e.g. in Fig. (\ref{SpectralIndex}), already the uncertainty in the measured value of the top-quark mass alone can shift the lower bound about $\pm 4$ GeV.}

\section{Frame Dependence of Quantum Corrections}
Despite the appealing nature of this unified scenario and the encouraging numerical results, there still remain conceptual problems. Some of these problems were addressed in \cite{weRG}, but probably the most fundamental and  conceptually difficult problem is associated with the frame dependence of quantum corrections. While it is rather easy to see that the two formulations in the JF and EF are \emph{mathematically} equivalent at the level of the classical action and the equations of motion, we are interested in the question whether this equivalence still holds at the quantum level. The problem is more general and arises in all theories with a non-minimal coupling, but it has in particular consequences for the non-minimal Higgs inflation model as the beta functions that determine the RG flow are derived by the effective action. Considering only the first quantum corrections, the question of frame dependence can be paraphrased as the question whether the following diagram does commute or not.
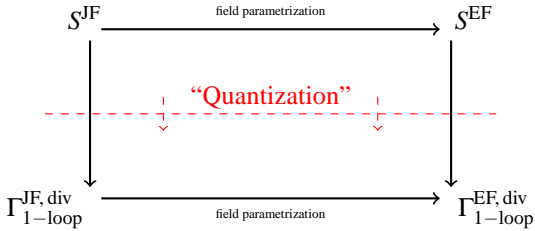
\begin{figure}[h!]
\begin{tikzpicture}[scale=0.75]
\draw[black,thick,->](0,0)--(6,0);
\draw[black,thick,->](0,-3)--(6,-3);
\draw[black,thick,<-](-0.2,-2.8)--(-0.20,-0.2);
\draw[black,thick,<-](6.2,-2.8)--(6.2,-0.2);
\draw[red,thin,dashed](-1,-1.5)--(7,-1.5);
\node[black](a) at(-0.3,0.2){$S^{\text{JF}}$};
\node[black](a) at(6.6,0.2){$S^{\text{EF}}$};
\node[black](a) at(-1,-3.2){$\Gamma^{\text{JF,\,div}}_{1-\text{loop}}$};
\node[black](a) at(7,-3.2){$\Gamma^{\text{EF,\,div}}_{1-\text{loop}}$};
\node[red](a) at(3,-1.2){``Quantization''};
\node[black](a) at(3,0.3){\tiny{field parametrization}};
\node[black](a) at(3,-3.3){\tiny{}field parametrization};
\draw[red,thin,dashed,->](1.1,-1.2)--(1.1,-1.8);
\draw[red,thin,dashed,->](4.9,-1.2)--(4.9,-1.8);
\end{tikzpicture}
\caption{Calculating quantum corrections in different parametrizations does not commute.}
\label{CommutingDiagram}
\end{figure}
In order to answer this question, we explicitly calculated the one-loop effective action in both parametrizations. Then we expressed the EF effective action in the JF parametrization and compared it with the JF effective action. We found that both results do not coincide \cite{weNew}.

\section{Origin of Frame Dependence and Vilkovisky's Idea}
In order to understand the origin of this frame dependence, we have to embed our cosmological model into the more general setup of field theory. Already in the mid-eighties, Gregori A. Vilkovisky proposed a solution to the problem of gauge and parametrization dependence of the effective action \cite{Vilk}. We will roughly sketch the main idea and then discuss the relevance for our cosmological problem. Let us adopt the condensed DeWitt notation, where discrete internal and space-time indices as well as the continuous space-time point are collected in one single index $i,j,...$. We can then describe a model by a single generalized field $\phi^{i}$ with the understanding that the Einstein summation convention is extended to include integration over space-time. When we think of the configuration space of fields ${\cal C}$ as differentiable manifold, we can think of $\phi$ as a point in ${\cal C}$. In this picture, different parametrizations $\phi^{i}$ and $\hat{\phi}^{i}$ that describe the same point simply correspond to different choices of coordinate systems in ${\cal C}$.

We would like to calculate the effective action and follow the standard procedure. We start with the Feynman path integral $Z[J]=\int{\cal D}\phi\,e^{i(S[\phi]+J_{i}\phi^{i})}$ which is a functional of the source $J$. The functional $W[J]$ that generates all \emph{connected} Greens functions is defined by $Z=e^{iW[J]}$. The \emph{mean field} is defined as $\langle\phi_{k}\rangle:=\frac{\delta W[J]}{\delta J^{k}}$. The effective action can then be obtained as functional Legendre transformation $\Gamma[\langle\phi\rangle]=W[J]-J_{k}\langle\phi^{k}\rangle$
and leads to the exact equation for the full effective action
\begin{align}
\text{e}^{i\,\Gamma[\langle\phi\rangle]}=\int{\cal D}\phi\,\text{e}^{i\,\left\{S[\phi]-\frac{\delta\Gamma[\langle\phi\rangle]}{\delta\langle\phi\rangle^{k}}\,\left(\phi^{k}-\langle\phi\rangle^{k}\right)\right\}}\label{EffActFull}
\end{align}
This equation can be solved iteratively in powers of $\hbar$ and yields at first order the
one-loop contribution
\begin{align}
\Gamma_{1-\text{loop}}=\frac{1}{2}\,\text{Tr}\,\text{ln}\,S_{,\, i j}\;.\label{EffActOneLoop}
\end{align}
Vilkovisky observed that the term $\left(\phi^{k}-\langle\phi\rangle^{k}\right)$ in (\ref{EffActFull}) is a \emph{coordinate difference} with respect to configuration space and has no \emph{geometrical} meaning.
He therefore proposed to replace it by a geometrically meaningful quantity. He borrowed the concept of the ``world function'' $\sigma$ that was originally introduced by Synge \cite{Synge} in the space-time context and lifted it up to the configuration space:
\begin{align}
2\,\sigma[\phi,\langle\phi\rangle]=(\text{geod. dist. between $\phi$ and $\langle\phi\rangle$})^2\;.
\end{align}
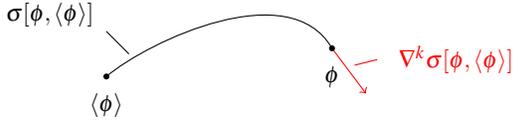
\begin{figure}[h!]
\begin{tikzpicture}[scale=1.5]
\draw (1,0) .. controls +(40:0.5cm) and +(120:0.75cm) .. (3,+0.25);
\draw[black](1,0.4)--(1.2,0.2);
\node[black](a) at(0.5,0.55){\small $\sigma[\phi,\langle\phi\rangle]$};
\node[black](a) at(1,-0.25){\small $\langle\phi\rangle$};
\node[black](a) at(3,0.0){\small $\phi$};
\draw[red,thin,->](3,0.25)--(3.3,-0.15);
\draw[red](3.2,0.1)--(3.4,0.15);
\node[red](a) at(4.1,0.15){\small $\nabla^{k}\sigma[\phi,\langle\phi\rangle]$};
\fill[color=black] (1,0) circle (0.75pt);
\fill[color=black] (3,+0.25) circle (0.75pt);
\end{tikzpicture}
\caption{Geodesic in ${\cal C}$, connecting the points $\phi$ and $\langle\phi\rangle$.}
\label{Geodesic}
\end{figure}
In order to render the formalism covariant, the coordinate difference is replaced by the geometrical meaningful covariant derivative of the world function
\begin{align}
\left(\phi^{k}-\langle\phi\rangle^{k}\right)\to\nabla^{k}\sigma[\phi,\langle\phi\rangle]\;.
\end{align}
At the one loop-level (\ref{EffActOneLoop}), the covariant reformulation leads to a replacement of the ``partial'' functional derivatives by covariant ones
\begin{align}
S_{,\,ij}\to\nabla_{i}\nabla_{j}S=S_{,\,ij}-\Gamma^{k}_{\;\;ij}\,S_{,\,k}\,.\label{CovDiffAction}
\end{align}
Vilkovisky proposed some physical assumptions that would fix the configuration space metric $G_{ij}$ and the configuration space connection $\Gamma^{k}_{\;\;ij}$ and would lead to a ``unique'' effective action. Regardless of whether we are willing to accept these assumptions, there are several conclusions - independent of these assumptions - that can be drawn from the covariant reformulation and help to shed some light onto the cosmological debate.

\section{The Cosmological Debate}
First of all, we notice that JF and EF are just two parametrizations among infinitely many others. It is also clear that the cosmological debate is just a very special case of the more general problem of parametrization dependence of the effective action. The analysis of the previous section shows why the off-shell results of the naive formulation of the effective action calculated in two different parametrizations do not coincide. Moreover, from (\ref{CovDiffAction}) we also see why the one-loop on-shell effective action is independent of the parametrization.\footnote{It is important to remember that in the derivation of (\ref{EffActOneLoop}) we have at no point made use of the equations of motion.} On-shell, the extra term in (\ref{CovDiffAction}) vanishes \emph{independently} of $\Gamma^{k}_{\;\;ij}$, which is in agreement with the on-shell theorems that prove the parametrization independence of the S-matrix.

Having detected the origin of the frame dependence as the lack of covariance of the \emph{mathematical} formalism, we can now turn to the discussion of the \emph{physical} consequences. In the aforementioned cosmological debate often \emph{physical} arguments are given in favour or against one or the other frame. A popular argument that should serve to support the JF as the ``real'' or ``physical'' frame is the claim that we measure real physical distances and time intervals with the Jordan frame metric.\footnote{What is missing in this argument is the fact that we can measure only \emph{dimensionless} quantities. In particular, this means that if we want to measure dimensionful quantities like Newton's constant $G_{\text{N}}$, this is only possible with respect to another dimensionful reference quantity $U$. This reference quantity $U$ is build from units like kilogram, meter, seconds etc. In the cosmological context of the Jordan frame the effective gravitational constant depends on $\varphi(\vec{x},t)$ . Since $\varphi$ is a scalar function that depends on space $\vec{x}$ and time $t$, it leads to a varying $G_{\text{N}}$. Consider for simplicity a homogeneous scalar field $\varphi(t)$ and imagine we would perform a Cavendish-type experiment at two different times $t$ and $t'$ in order to determine whether $G_{\text{N}}$ is time dependent or not. Suppose that we would find $G_{\text{N}}(t)/U(t)-G_{\text{N}}(t')/U(t')\neq 0$. How can we say whether the gravitational constant $G$ has changed or whether the units (or both) have changed?  We simply cannot! All we can say is that the \emph{ratio} $G_{\text{N}}/U$ has changed with time. Among the infinitely many parametrizations, JF and EF are distinguished in the sense that they correspond to the extreme cases where either $G_{\text{N}}$ or $U$ are constant. In the JF $G_{\text{N}}$ is varying, while $U$ is constant and vice versa in the EF.}
However, from the above considerations it should be clear once and for all that \emph{there is no distinguished physical frame}. It is pointless to talk about the \emph{physical} meaning of one or the other frame in the same sense as it is meaningless to ask whether it is more physical to describe a mechanical system in spherical or Cartesian coordinates. If we describe the same physical theory, the result will of course not depend on the parametrization as long as the formalism is covariant. Some parametrizations may be preferred in the sense that they are more adapted to the symmetries of the underlying theory, but Nature certainly does not care about how we parametrize her.

\begin{theacknowledgments}
We thank A. O. Barvinsky, C. Kiefer and A. A. Starobinsky for numerous fruitful discussions. C. S. was partially supported by the ERC Grant 277570-DIGT. A. K. acknowledges
support by the RFBR grant 11-02-00643.
\end{theacknowledgments}
\bibliographystyle{aipproc}   

\end{document}